\documentclass[11pt]{article}
\usepackage{moriond,epsfig,epsf}

\newcommand{\dslash}{\partial\hspace{-.075in}/}

\newcommand{\lsim}{\mathrel{\lower4pt\hbox{$\sim$}}
\hskip-12.5pt\raise1.6pt\hbox{$<$}\;}
\newcommand{\gsim}{\mathrel{\lower4pt\hbox{$\sim$}}
\hskip-12.5pt\raise1.6pt\hbox{$>$}\;}
\def\be{\begin{equation}}
\def\ee{\end{equation}}
\def\bea{\begin{eqnarray}}
\def\eea{\end{eqnarray}}

\begin{document}
\vspace*{4cm}
\title{ENERGETICS OF NEUTRINOS IN  NEUTRON STARS}

\author{Ken Kiers$^1$, Michel  
H.G. Tytgat$^{2,}$ \footnote{Talk presented 
by M.T. at the XXXIVnd Rencontres de Moriond: ELECTROWEAK INTERACTIONS AND UNIFIED THEORIES} }

\address{\vspace{0.5cm}
1. Physics Department, Taylor University\\
 236 West Reade Ave.
                                     Upland, IN 46989
                                     USA\\
\vspace{0.2cm}
2. Service de Physique Th\'eorique, CP225\\
Universit\'e Libre de Bruxelles, Bld du Triomphe, 1050 Bruxelles, Belgium}

\maketitle

\abstracts{I review our proof  that
  long range forces  induced by  the exchange  of massless neutrino-antineutrino pairs
 do not affect the stability of neutron stars.
}

\newpage
\section{Introduction}
\label{subsec:prod}

\begin{figure}[thb]
\centerline{\epsfig{figure=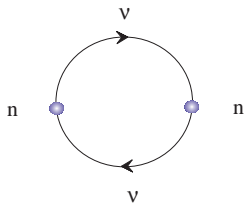,height=2.5cm}}
\caption{Exchange of a neutrino-antineutrino pair between two neutrons (n) at rest.}
\label{figone}
\end{figure}

Consider two test neutrons at rest, separated by the distance $d \gg 1/M_Z$. 
If neutrinos are massless, they induce  (see Fig.~\ref{figone}) an effective  long range 
interaction between the two neutrons which, on dimensional grounds, must fall like
$$
V_{(2)}(d) \sim {G_F^2 \over d^5}
$$
($G_F \approx 10^{-5} GeV^{-2}$ is the Fermi coupling.)~\cite{feinberg}. 
This interaction is very  weak: at short distances, $d \lsim 10^{-8} cm$, it is dominated by 
the direct exchange of a $Z$ boson; at large distances, $d \gsim 10^{-8} cm$,  the gravitational interaction (!) between the two test
neutrons is dominant.
Similarly, in the presence of $N$ neutrons (taken to be at rest, to simplify),  long range, multibody interactions
are induced by the exchange of a neutrino-antineutrino pair
between any  subset of $k$  neutrons, $k \leq N$.
\footnote{One-loop diagrams with an odd number of neutron insertions
vanish.} Consider then  the contribution of these interactions to the
total potential energy of the $N$ neutrons. As each coupling to a neutron  brings a factor of $G_F$,
 the  k-body contribution gets smaller as $k$ increases. However,  each contributions has to be added 
with the adequate combinatoric factor, that, at least superficially,  is proportional 
to the number of ways to take $k$ neutrons out of $N$, $\propto N^k$  for large $N$. 
\begin{figure}[thb]
\centerline{\epsfig{figure=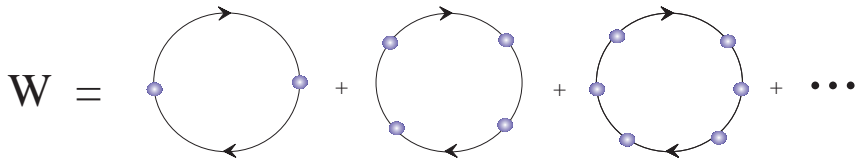,height=2cm}}
\label{figtwo}
\caption{Expansion of the self-energy in k-body neutrino-antineutrino exchange.}
\end{figure}
To be specific, consider a neutron star, that contains $N \approx 10^{57}$ neutrons -- corresponding
to  about $1.4$ solar mass-- within a volume of radius $R \approx 10 km$, Fig.[3].
The self-energy of the neutron star from neutrino exchange is naively represented by the 
serie of Fig.[2], where the dots now mean  insertions of the  neutron density, $\rho_N$. The term
with $k$ insertions of the neutron density 
scales approximately as
\equation
        W^{(k)} \sim \frac{C_k}{R}\left(\frac{G_F N}{R^2}\right)^k, 
\endequation
where $C_k$ is
a dimensionless numerical coefficient.  The dimensionless parameter that governs the expansion
is thus  $G_F N/R^2 \sim G_F \rho_n R$. For a neutron star, this parameter is  ${\cal O}(10^{12})$ (!).

Attempts to a direct summation of this series (truncated
after $N$ terms) have  yielded  enormous values for the interaction self-energy~\cite{fb}, which  
led some  authors
to claim that neutrinos must have a mass of at least $0.4$ eV
in order to allow neutron stars to exist~\cite{fb,fb2,fischxxx}.\footnote{This problem has an interesting story.
It was first raised by R. Feynman in his Caltech lectures on gravity~\cite{feynman}. He asked himself whether long range neutrino
forces could mimic gravity, but  ruled it out because the equivalence principle was not satisfied. He  noticed  that a large
quantity of matter would render the expansion in k-body interactions divergent. Hartle addressed this problem in 
a cosmological context~\cite{hartle}. He noticed that the large scale repartition of matter in the Universe
is such that the perturbative expansion is actually well-defined and leads to totally negligeable effects.
 He also remarked that the long range neutrino interactions have no effect in presence of an homogeneous distribution of matter.
 Fischbach later recognized that the 
is expansion parameter is  $\gg 1$ inside of  a neutron star~\cite{fb}.}
\begin{figure}[thb]
\centerline{\epsfig{figure=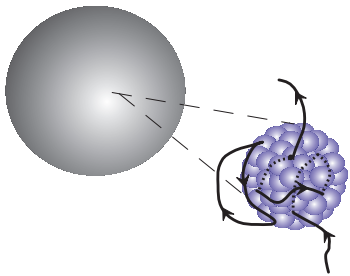,height=4cm}}
\caption{Exchange of neutrino-antineutrino pairs between the neutrons of a neutron star.}
\label{figthree}
\end{figure}
Our approach to this problem is different. Following Abada {\em et al}, we claim that the series 
represented in Fig.[2] is meaningless if $G_F \rho_N R \gsim 1$, and must be resummed 
in a non-perturbative way to 
get a sensible result~\cite{abada,kt}. 
We have performed this
resummation and find that the apparent infrared divergence
is an artifact of the expansion and that the energy is
finite and well-behaved. This implies in particular that there is no lower bound on
the mass of the neutrinos.  Along the way we have encountered some
interesting physics. In particular, we show that the
ground state of the system has a non-zero neutrino
charge -- a result which was previously anticipated by Loeb~\cite{loeb}.
In our simple model for the density of the neutron star it is
straightforward to calculate both the energy and the neutrino
number of a neutron star.
Let us
mention that several groups  have examined
various aspects of this 
problem~\cite{abada,smirnov1,recone,rectwo,penexxx,arafune}.
In particular, Arafune and Mimura~\cite{arafune}
have confirmed our asymptotic result using an analytical approximation.

\section{The neutrino ground state in a neutron star}

We want to compute the contribution to the self-energy of the neutron star due to long range neutrino forces, or equivalently, the shift
in the neutrino ground state energy in presence of a neutron  star. The latter can be defined in terms of the neutrino Hamiltonian 
$H_{(0)}$ in the presence (absence) of
the star~\cite{julian}:
\begin{equation}
\label{shift}
W = \langle \hat 0 \vert  H \vert \hat 0 \rangle - \langle 
 0 \vert  H_0 \vert  0 \rangle .
\end{equation}
Here $\vert \hat 0 \rangle$ denotes the neutrino
ground state in  presence of the star, while $\vert 0 \rangle$
denotes the usual neutrino vacuum state.  As we have already
alluded, the state $\vert \hat 0 \rangle$ contains in general a 
non-zero neutrino number (i.e., it is ``charged'').
Note that the expression in Eq.~(\ref{shift}) is a formal, 
ultraviolet (UV) divergent quantity which needs to be renormalized.  
This renormalization may be done using the usual techniques of quantum field theory.

In order to proceed, it is convenient to introduce a low energy 
effective Lagrangian for the neutrino field.  After integrating
out all of the other particles in the theory, one 
obtains~\cite{mann,abada,ken2}
\begin{equation}
\label{efflag}
{\cal L}_{\rm eff} = 
  {\psi}_{L} \left[i\dslash 
        + \alpha \gamma^0\right]\,\psi_{L}
\end{equation}
where $\psi_L = \frac{1}{2}(1-\gamma_5) \psi$, and where
\begin{equation}
\alpha(\vec{x}) = G_F\rho_n(\vec{x})/ \sqrt{2}\, \sim 20\,\,{\rm eV}
\end{equation}
is the  electroweak  potential 
induced by the finite neutron density 
(typically $\rho_n \approx 0.4\,\, \mbox{\rm fm}^{-3}$ 
in a  neutron star).  This potential is
identical to the one which is usually considered 
in the well-known Mikheyev-Smirnov-Wolfenstein (MSW) effect~\cite{msw}.
The potential term in Eq.(\ref{efflag}), which is attractive for neutrinos and repulsive for antineutrinos, resembles
a position-dependent chemical potential, and it is maybe not  surprising that the ground state of the system can 
carry a non-zero neutrino number.\footnote{This conclusion could be premature. For instance, in $1+1$ dimension,  such an external  potential 
has absolutely  no physical effect
(if the neutrinos are massless)~\cite{recone}.}
\footnote{From the effective Lagrangian point of view, it is manifest  that an homogeneous distribution 
of neutrons, $\rho_n = $ const. has no net effect: it can  simply be gauged away. (See Hartle's comment in footnote {\em b}.)}

The Schwinger-Dyson 
 expansion for $W$ in terms of the potential $\alpha(\vec{x})$
and the neutrino propagator $G_0(\vec{x},\vec{x}^\prime;\omega)$ gives~\cite{kt}
\begin{eqnarray}
\label{exp2}
        W &  = & {1\over 2 \pi i} \sum_{k = 1}^\infty \, {1 \over k}\,
 \int_C d\omega\, 
 \mbox{\rm Tr}_{\bf x} \left  [{\alpha  G_0(\omega)}\right ]^k \\
        & \equiv & \sum_k W^{(k)} .
        \label{eq:pertexp}
\end{eqnarray}
This series corresponds precisely 
to the one  which is represented diagrammatically in Fig.[2].
A useful aspect of the perturbative expansion
is that it  isolates
the UV divergence in $W$.  In particular, the only UV divergent term in Eq.(\ref{eq:pertexp}) is that with $k=2$ which
is related to the vacuum polarization of the $Z$ boson in the
complete theory.  While the terms with $k\geq 4$ are separately UV finite,
their sum is ill-defined
for $\alpha R \gsim 1$.  The non-perturbative
resummation of these terms is the main goal of our calculation.

\subsection{The 2-body contribution.}

We begin by an estimate of the two-body contribution to the self-energy of a neutron star, $W_2$. From Eq.(\ref{efflag})
this contribution  is UV  divergent, $W_2 \sim \Lambda^2$, with $\Lambda$  some $UV$ cut-off . The procedure to 
compute this term is in principle well-defined. The diagram can be computed and renormalized both in the effective theory and in the underlying
theory -- {\em i.e.} the Standard Model -- and the respective results must be matched at some scale. A natural scale for matching is 
the inverse size of the neutrons, 
$\Lambda \sim 1 GeV$. From the effective Lagrangian point of view, this
implies that the coarse grained structure of the star  is taken into account. At larger distances, $d \gg GeV^{-1}$
it is however sufficient to consider a smooth,  continuous distribution of neutrons. This simplification is helpful
when we  come to the evaluation of the higher order, $UV$ finite,  multibody contributions.

A rough  estimate for $W_2$ is~\cite{fb,abada,kt,penexxx} 
\begin{equation}
\label{est}
  W_{2} \sim   +\, \alpha^2 \, \Lambda^2 \, R^3 
\end{equation}
The two-body interactions give a contribution  $\propto R^3$,
and  proportional to the number of neutron pairs in the star, $\propto \rho^2 \sim N^2$.
Also, because the two-body interaction is repulsive, this contribution  is positive.
For a neutron star, this estimate gives
a negligeable contribution,
\begin{equation}
  W_{2} \sim  10^{16} kg \ll 1.4 \, M_\odot \sim 10^{31} kg.
\end{equation}

\subsection{The $k$-body contributions, $k \geq 4$.}

The terms in the Schwinger-Dyson expansion with $k\geq 4$ are UV finite. However, there
are sensitive to the size of the star,  
$\propto (\alpha \,
R)^k$. For $\alpha \,
R > {\cal O}(1)$ these terms must be resummed in a non-perturbative way. We begin by making  some useful
simplifications. First, to study long distance, $UV$ finite effects, we can forget the neutrons
and only consider their mean field effect: $\alpha(x)$ is taken to be a smooth
function of $\bf x$. Furthermore, we approximate the shape of the star by a spherical square well potential, 
with depth $\alpha \approx 20 eV$ and radius $R$. For $k \geq 4$, the only dimensionless parameter is 
 $\alpha  R$. If $\alpha R \lsim 1$  the Schwinger-Dyson expansion
is well-defined (the "weak coupling regime''), while it must be resummed if $\alpha R \gsim 1$ 
(the "strong coupling regime''). 
If we want to understand the transition
from the "weak coupling" to the "strong coupling" regime,  we only have consider systems of rather small size: 
{\em e.g.} for fixed $\alpha \approx 20 eV$, $0.1 \leq \alpha R \leq 100$. The extrapolation to $R \sim 10 km$ will be trivial.

To resum the $k \geq 4$ terms, we use the following 
expression --adapted from Schwinger~\cite{julian}-- for the shift of the neutrino vacuum energy in presence of the external electroweak potential:
\begin{equation}
W = \frac{1}{2\pi} \sum_{l=0}^{\infty} (2l+2) \int_{0}^\infty
                d\omega \; \left[\delta_l(\omega)+\delta_l(-\omega)
                \right] . \label{wformal}
\end{equation}
where   $\delta_l(\pm\omega)$ is the scattering phase shift of an incident neutrino (antineutrino)
of energy $(-)\omega$  and $l$ labels the orbital angular momentum.
The factor $(2l+2)\equiv (2 j +1)$ is the degeneracy factor for a given
energy $\pm \omega$ and total angular momentum $j$.

\noindent
The details of our calculations can be found in~\cite{kt}. Here, I only summarize the main results. 
\begin{description}
\item{-} 
Eq.~(\ref{wformal}) is a  formal $UV$-divergent expression 
which need to be renormalized. We achieved this by substracting the leading $W_2={\cal O}(\alpha^2)$
term in the Born expansion of the phase shifts. The "renormalized" term $W_2$ is then like in Eq.(\ref{est}).
\item{-} If the neutrino are massless, there are no bound states, only resonances.
This is because a neutrino must be able to flip chirality $\nu_L \rightarrow \nu_R$ in order to form a bound state.
Resonances exists because  no chirality flip is necessary to bend the trajectory of a massless neutrino. (See Loeb~\cite{loeb}.) 
These resonances become extremely narrow -- {\em i.e.} long lived -- as $R$ increases.
\item{-} The physical implication  of these resonances  is that  a non-zero neutrino charge can be ``confined" within the potential/neutron star.
Intuitively, the external electroweak potential induced by the finite neutron density polarizes the neutrino vacuum. If the external
field is strong enough, {\em i.e.} $\alpha R \gsim 1$, the neutrino vacuum is unstable toward the creation of neutrino-antineutrino
pairs. In an open system, like a neutron star,  the 
antineutrinos fly  away to spatial infinity
while the neutrinos effectively ``screen'' the external electroweak potential. This effect is obviously
non-perturbative in the external potential.
\begin{figure}[thb]
\centerline{\epsfig{figure=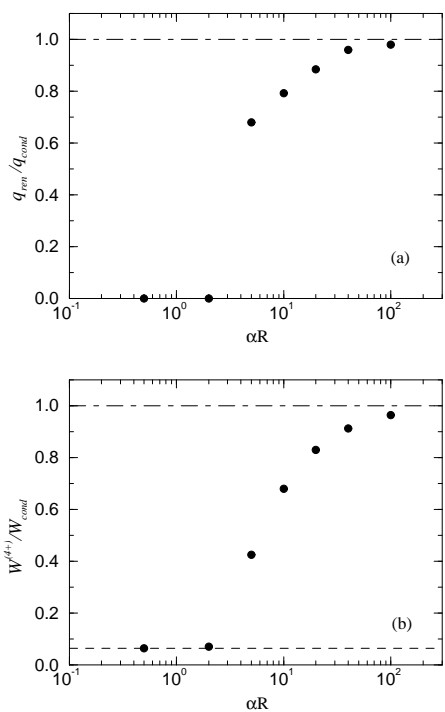,height=10cm}}
\caption{Renormalized exact charge, $Q/Q_\nu$, and self-energy $\vert (W - W_2)/W_\nu\vert$ as a function of $\alpha R$.} 
\label{figthree}
\end{figure}
\item{-} As $R$ increases, we observe a first order phase transition at $\alpha R \sim 1$ 
between the ``weak coupling'' and the ``strong coupling'' regimes,  after which the self-energy
rapidly reaches an asymptotic behaviour, Fig.[4]. The calculations show that 
this asymptotic value coincides with the thermodynamic limit of a ideal Fermi gas of 
massless neutrinos  with chemical potential $\alpha$, trapped within a volume $V = 4/3 \pi R^3$:
\begin{eqnarray}
W - W_{2}  &\approx&  W_\nu = - {\alpha^4 R^3\over 18 \pi}\nonumber\\
Q &\approx& Q_\nu = 2 {(\alpha R)^3\over 9 \pi}
\end{eqnarray}
This behaviour for very large $\alpha R$  was 
anticipated by Abada {\em et al}~\cite{abada} and confirmed by Arafune and Mimura~\cite{arafune}.
\item{-} Because there is no other dimensionless parameter in the theory than  $\alpha R$, 
it is trivial to extrapolate these results to a realistic neutron star, $R \sim 10 km$. 
There are about
  \begin{equation}
 Q_\nu \approx 10^{36}   
  \end{equation}
low energy, $\omega \sim eV$, neutrinos trapped in a neutron star.  
Expressed in kilograms, they contribute about
\begin{equation}
W - W_2 \approx - 30 kg
\end{equation}
(for each generation of massless neutrinos) to the mass of the star, to be compared to a total mass of $10^{31} kg$.
\item{-} This effect is extremely small, probably of no observable consequence. Incidentally, this proves that massless neutrinos
can not affect the gravitational stability of a neutron star.
\end{description}

\section{Conclusions}

The physics underlying the resummation problem is non-perturbative. 
The neutrons are the source of a neutral electroweak effective potential that polarizes the
neutrino vacuum. When the system is large/dense enough (the "strong coupling" regime), like in a neutron star,
 the energy of the system is
 lowered by spontaneous creation of neutrino-antineutrino pairs.
In an open system, again  like a neutron star,  the
 ground state 
contains a "neutrino condensate". These results show that 
there is no ``mysterious'' long range neutrino force at work in a neutron star~\cite{raffelt}.


\section*{References}
\begin{thebibliography}{99}
%

\bibitem{feinberg}
G. Feinberg and J. Sucher, Phys. Rev. {\bf 166}, 1638 (1968);
G. Feinberg and J. Sucher and C.-K. Au,  Phys. Rept.
{\bf 180}, 83 (1989);
S.D.H. Hsu, P. Sikivie, Phys.Rev. {\bf D49}, 4951 (1994);
J.A. Grifols, E. Masso and R. Toldra, Phys.Lett. {\bf B389}, 563
(1996).
%
\bibitem{feynman}
R.P. Feynman, F.B. Morinigo and W.G. Wagner, Feynman lectures on 
Gravitation, Addison-Wesley, Readings, MA (1995).
%
%
\bibitem{hartle}
J.B. Hartle, Phys. Rev. {\bf D1}, 394 (1970).
%
\bibitem{fb}
E. Fischbach,  Ann. of Phys. {\bf 247}, 213 (1996).
%
\bibitem{fb2}
B. Woodahl, M. Parry, S.-J. Tu and E. Fischbach, hep-ph/9709334.
%
\bibitem{fischxxx}
E. Fischbach and B. Woodahl, hep-ph/9801387.
%
\bibitem{kt}
K. Kiers and M.H.G. Tytgat, Phys. Rev. D {\bf 57}, 5970 (1998).
%


\bibitem{abada}
As. Abada, M.B. Gavela and O. P\`ene, Phys. Lett. B {\bf 387}, 315
(1996).
%
 
\bibitem{loeb} 
A. Loeb, Phys. Rev. Lett. {\bf 64} (1990) 115.
%
\bibitem{smirnov1}
A. Y. Smirnov and F. Vissani, hep-ph/9604443.
%
\bibitem{recone}
 As. Abada, O. P\`ene and J. Rodriguez-Quintero, Phys.Lett. {\bf B423}, 355 (1998).
%
\bibitem{rectwo}
M. Kachelriess, hep-ph/9712363.
%
\bibitem{penexxx}
As. Abada, O. P\`ene and J. Rodriguez-Quintero, Phys.Rev. {\bf D58}, 073001 (1998);
{\em ibid} {\bf D59}, 077302 (1999). 
\bibitem{arafune}
J. Arafune and Y. Mimura, Prog.Theor.Phys. {\bf 100}, 1083 (1998). 
%
\bibitem{julian}
J. Schwinger, Phys. Rev. {\bf 94}, 1362 (1954).
%
\bibitem{mann}
P.D. Mannheim, Phys. Rev. {\bf D 37}, 1935 (1988).
%
\bibitem{ken2}
K. Kiers and N. Weiss, Phys. Rev. {\bf D56}, 5776 (1997).
%
\bibitem{msw} L. Wolfenstein, Phys. Rev. D {\bf 17} (1978) 2369;
        Phys. Rev. D {\bf 20} (1979) 2634; \\
        S.P. Mikheyev and A.Yu. Smirnov, Yad. Fiz. {\bf 42} (1985)
        1441 [Sov. J. Nucl. Phys. {\bf 42} (1985) 913]; Il Nuovo 
        Cimento C {\bf 9} (1986) 17.
%
\bibitem{raffelt}
G.G. Raffelt,  hep-ph/9902271; hep-ph/9903472.
\end{thebibliography}
\end{document}